\documentclass[preprint,10pt]{elsarticle}

\usepackage{multirow}
\usepackage{epsfig}
\usepackage{graphicx}
\usepackage{subfigure}
\usepackage{svg}
\usepackage[colorlinks,linkcolor=red,anchorcolor=blue,citecolor=green]{hyperref}
\usepackage{siunitx}
\usepackage{amsmath}
\usepackage[utf8]{inputenc}
\usepackage{bm}
\usepackage{amssymb}
\journal{Science China Technological Sciences}

\begin{document}

\begin{frontmatter}



  \title{Scaling Laws Governing the Elastic Properties of 3D-Graphenes}


  \author[inst1,inst2]{Ming Li}
  \author[inst3]{Guo Lu}
  \author[inst1,inst2]{Haodong Yu}
  \author[inst4]{Menglei Li*}
  \author[inst1,inst2]{Fawei Zheng*}
  \affiliation[inst1]{organization={Centre for Quantum Physics, Key Laboratory of Advanced Optoelectronic Quantum Architecture and Measurement (MOE), School of Physics, Beijing Institute of Technology},
    city={Beijing},
    postcode={100081},
    country={China}}
  \affiliation[inst2]{organization={Beijing Key Lab of Nanophotonics $\&$ Ultrafine Optoelectronic Systems, School of Physics, Beijing Institute of Technology},
    city={Beijing},
    postcode={100081},
    country={China}}
  \affiliation[inst3]{organization={LCP, Institute of Applied Physics and Computational Mathematics},
    city={Beijing},
    postcode={100088},
    country={China}}
  \affiliation[inst4]{organization={Department of Physics, Capital Normal University},
    city={Beijing},
    postcode={100048},
    country={China}}

  \begin{abstract}
  	In this study, we have comprehensively investigated the scaling law for elastic properties of three-dimensional honeycomb-like graphenes (3D-graphenes) using hybrid neural network potential based molecular dynamics simulations and theoretical analyses. The elastic constants as functions of honeycomb hole size, denoted by the graphene wall length $L$, were provided. All five independent elastic constants in the large $L$ limit are proportional to $L^{-1}$. The associated coefficients are combinations of two-dimensional graphene's elastic constants. High-order terms including $L^{-2}$ and $L^{-3}$ emerge for finite $L$ values. They have three origins, the distorted areas close to the joint lines of 3D-graphenes, the variation of solid angles between graphene plates, and the bending distortion of graphene plates. Significantly, the chirality becomes essential with the decreasing of $L$, because the joint line structures are different between the armchair and zigzag type 3D-graphenes. Our findings provide insights into the elastic properties of graphene-based superstructures and can be used for further studies on graphene-based materials.
  \end{abstract}





\end{frontmatter}

  \section{Introduction} \label{sec:introduction}

  Since its first successful isolation\cite{Novoselov2004Electric} in 2004, graphene has attracted significant research interest in the fields of physics, chemistry, and material science. Graphene possesses various fascinating  properties \cite{Sun20203DGraphene}, including the large specific surface area ($\sim2630  \si{\metre}^2/\si{\gram}$) \cite{li2014ultrahigh}, strong chemical stability \cite{suzuki2017chemical}, high mechanical strength (with Young's modulus of $\sim340$ N/m) \cite{lee2008measurement}, and excellent thermal conductivity ($\sim5000  \si{\watt}/\si{\milli\kelvin}$ at room temperature) \cite{Balandin2008Superior}. However, in practical applications, the two-dimensional (2D) graphene layers tend to restack and lose those outstanding properties, resulting in the performances of the graphene-based devices, for instance supercapacitors with graphene as electrode materials, far below the theoretical expectations \cite{Liu2010Graphene}. To overcome this issue, one approach is to arrange the 2D layers into spatially well-organized three-dimensional (3D) configurations that maintain the excellent properties of 2D graphene without the concern of restacking \cite{Sun20203DGraphene}. In recent years, numerous graphene-based 3D architectures, including macroscopic graphene hydrogels, foams, sponges, and microscopic flower- or honeycomb-like graphene, have been synthesized using a diversity of methods such as graphene oxide by chemical reduction, hydrocarbons via chemical vapor deposition, and inorganic carbon compounds based on a series of alkali-metal-involved chemical reactions \cite{RN2431,RN2430,RN2433,RN2432,RN2449}. In these methods, 3D graphene structures are successfully fabricated mainly through combining the 2D graphene layers into strongly bonded 3D graphene superstructures. One well known superstructure is the 3D honeycomb-like graphene structure (In this paper we will use the word '3D-graphene' to refer specifically to this structure), which was initially conceived by Karfunkel \emph{et al}. more than thirty years ago \cite{RN2450}. Recently, this honeycomb-like 3D-graphene has been successfully grown in experiment by Krainyukova \emph{et al} \cite{RN2451}. Thanks to the strong valence bonds between graphene plates, 3D-graphene has a wider range of applications \cite{RN2451,RN2452,RN2453}.

  For practical applications of 3D-graphene, a comprehensive understanding of the elastic properties is essential. A number of studies have been carried out in this regard. Pang \emph{et al.} \cite{RN2454} investigated the behaviors of 3D-graphene when stretched in various directions and found that the structure has high specific strength, which is tunable via the cell size. Zhang \emph{et al.} \cite{RN2455} used  molecular dynamics (MD) simulations and first-principles calculations to study 3D-graphene with different 'wall-chirality' and sizes, discovering that the Young's modulus depends solely on the hinge density while the failure strain is influenced by the lattice size and geometrical regularity. Meng \emph{et al.} \cite{RN2456} employed numerical simulations and theoretical modelings to systematically investigate the behaviors of 3D-graphene upon out-of-plane compression, revealing that two critical deformation types, i.e. elastic mechanical instability and inelastic structural collapse, depend on the cell size and local atomic bonding configurations at cell junctions.

  Although there has been some progress in elucidating the mechanical properties of 3D-graphenes, a comprehensive understanding of their elastic properties remains to be achieved. In this work, we have performed a systematic investigation of the elastic properties of 3D-graphenes by using a combination of MD simulations and theoretical analysis. We have employed MD simulations to calculate the elastic constants for 3D-graphenes with various cell sizes and chirality. We have found that all five independent elastic constants follow $L^{-1}$ scaling law, where $L$ is the length of the graphene wall. Minor deviations described by $L^{-2}$ and $L^{-3}$ terms  were  proposed to describe the contributions from the areas near joint lines and bending distortions. The high-order terms show chirality-dependent behaviors, while the $L^{-1}$ term is solely determined by the elastic properties of graphene, independent of the chirality of 3D-graphenes.

  \section{Atomic Models and Computational Methods} \label{sec:Computational Models and Methods}

\begin{figure}[htbp]
    \centering
    \includegraphics[width=0.8\textwidth]{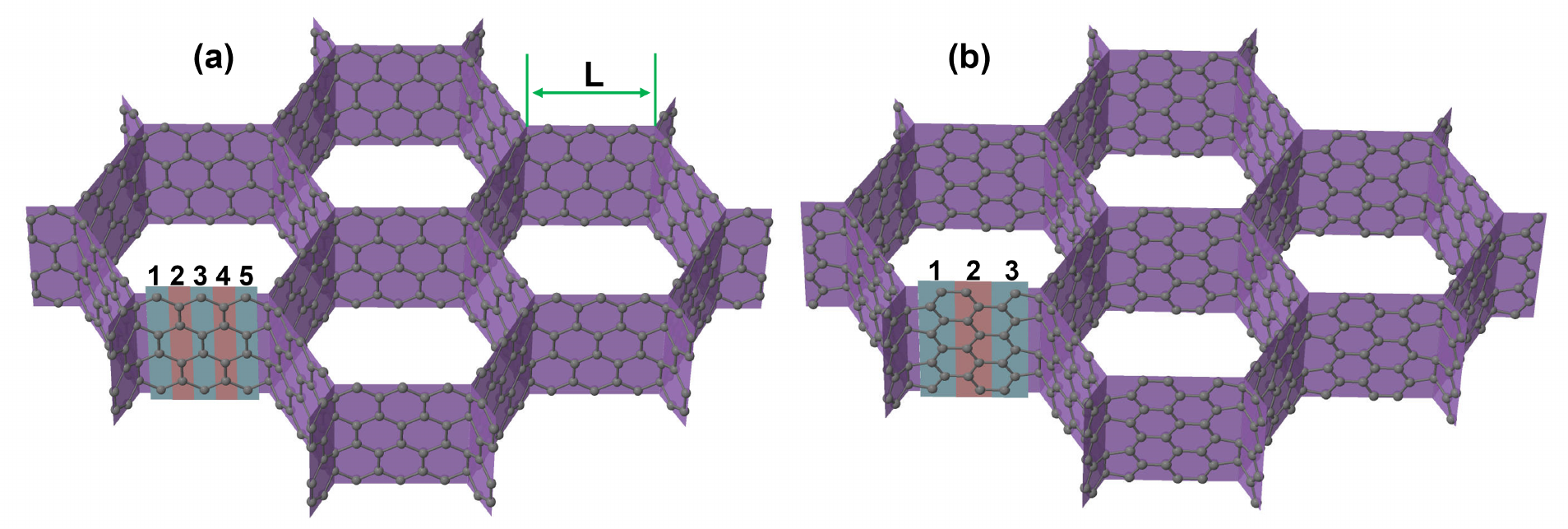}
    \caption{The atomic structure of 3D honeycomb-like graphene. (a) A slice of Z3DG-5, which has zigzag edges across the $\bm{a}$-$\bm{b}$ plane, the length of graphene 'wall' was shown as $L$. (b) A slice of A3DG-3, which has armchair edges across the $\bm{a}$-$\bm{b}$ plane.}
    \label{3DG_structure}
  \end{figure}

  3D-graphenes have hexagonal crystal structures with two independent lattice constants, namely $a=|\bm{a}|=|\bm{b}|$ and $c=|\bm{c}|$, where $\bm{a}$, $\bm{b}$, and $\bm{c}$ are the three lattice basis vectors. The lattice parameter $c$ is nearly a constant for 3D-graphenes with different size. On the other hand, the lattice constant $a$ is dependent on the length of the honeycomb 'wall'. In this study, we numerically studied 3D-graphenes with $a$ smaller than 200 \AA. As shown in Fig. \ref{3DG_structure}(a-b), each unit cell of 3D-graphene  consists of three 'walls', which are three graphene plates. If we cut a 3D-graphene along $\bm{a}$-$\bm{b}$ plane, each graphene plate will exhibit either zigzag or armchair edges. Because all three graphene plates possess the same chirality, we label a 3D-graphene with zigzag edges as Z3DG (Fig. \ref{3DG_structure}(a)), and a 3D-graphene with armchair edges as A3DG (Fig. \ref{3DG_structure}(b)). To distinguish the structures with different lengths of graphene plates, a structural index $m$ is defined to represent the number of carbon atomic chains in each graphene plate. Z3DG with $m$ columns of carbon chains is denoted as Z3DG-$m$, such as Z3DG-5 in Fig. \ref{3DG_structure}(a). Similarly, A3DG with $m$ columns of carbon chains is labeled as A3DG-$m$, such as A3DG-3 in  Fig. \ref{3DG_structure}(b).

  In this study, the mechanical properties of 3D-graphenes were calculated through classical MD simulations utilizing the General Utility Lattice Program (GULP) \cite{gale1997gulp,gale2003general}. A hybrid neural network potential \cite{wen2019hybrid} for graphene system obtained through the OpenKIM \cite{tadmor2011potential} interatomic potential repository was used in our simulations. It employs a neural network to describe the short-range interactions and an analytical term to model the long-range dispersion for multilayer graphene. This potential has shown a high accuracy in reproducing results calculated from density functional theory for structural, energetic, and elastic properties.
  It enables accurate simulations of 3D-graphene at large scale, and makes the present work on scaling law possible.

  The elastic properties of a solid are described by the generalized Hooke's Law
  \begin{equation} \label{eq.2}
    \sigma_{ij}=\sum_{k=1}^{3}\sum_{l=1}^{3}{C_{ijkl}\cdot e_{kl}}
  \end{equation}
  where $\sigma_{ij}$, $e_{kl}$, and $C_{ijkl}$ denote the elements of the symmetric stress, strain, and the elastic tensor, respectively \cite{Raabe1998Computational}. The elastic tensor can be conveniently written in the following form using the Voigt notation \cite{Raabe1998Computational}:
  \begin{equation} \label{eq.3}
    \sigma_i=\sum_{j=1}^{6}C_{ij}\cdot e_j
  \end{equation}

  Based on the symmetries of the hexagonal lattice, the elastic tensor can be written as \cite{Raabe1998Computational}:
  \begin{equation}\label{C_hex}
    \bm{C}=
    \left(
    \begin{matrix}
        C_{11} & C_{12} & C_{13} & 0      & 0      & 0                                     \\
        C_{12} & C_{11} & C_{13} & 0      & 0      & 0                                     \\
        C_{13} & C_{13} & C_{33} & 0      & 0      & 0                                     \\
        0      & 0      & 0      & C_{44} & 0      & 0                                     \\
        0      & 0      & 0      & 0      & C_{44} & 0                                     \\
        0      & 0      & 0      & 0      & 0      & \frac{1}{2}\left(C_{11}-C_{12}\right) \\
      \end{matrix}
    \right),
  \end{equation}
  where the only five independent constants are $C_{11}$, $C_{12}$, $C_{13}$, $C_{33}$, and $C_{44}$.

  The elastic tensor for graphene, as a 2D hexagonal lattice, can be directly obtained by removing the 3rd, 4th, and 5th rows and columns of the matrix in Eq. \ref{C_hex}
  \begin{equation}\label{C_2D}
    \bm{C}^{g}=
    \left(
    \begin{matrix}
        C_{11}^{g} & C_{12}^{g} & 0                                             \\
        C_{12}^{g} & C_{11}^{g} & 0                                             \\
        0          & 0          & \frac{1}{2}\left(C_{11}^{g}-C_{12}^{g}\right) \\
      \end{matrix}
    \right).
  \end{equation}
  It has two independent components $C_{11}^g$ and $C_{12}^g$. Then the Poisson's ratio and Young's modulus can be written as:
  \begin{equation}
    \begin{aligned}
      \gamma & ={C_{12}^g}/{C_{11}^g} \\
      C_Y^g  & =C_{11}^g(1-\gamma^2)
    \end{aligned}
  \end{equation}

  In our calculations, the two independent elastic components of the graphene are $C_{11}^g=341.3$ N/m and $C_{12}^g=83.2$ N/m. Then the Poisson ratio is $\gamma=0.24$, which agrees with experiment results 0.19$\sim$0.36 \cite{politano2015probing}. The Young's modulus is 321.0 N/m, which also agrees well with experiment results 310$\sim$342 N/m \cite{politano2015probing} and 340$\pm$50 N/m \cite{lee2008measurement}.

  \section{Results and Analysis}

  As we know, all the carbon atoms in graphene are sp$^2$ hybridized, and that in diamond are sp$^3$ hybridized. Each carbon atom in graphene (diamond) has three (four) chemical bonds. In the case of 3D-graphene, most of the carbon atoms are sp$^2$ hybridized, while the carbon atoms at the joint lines are sp$^3$ hybridized.  For a A3DG-$m$ or Z3DG-$m$ with large index $m$, most of the carbon atoms are in the graphene plates with sp$^2$ hybridization. On the other hand, when the index $m$ is small, a considerable portion of carbon atoms are sp$^3$ hybridized. The ratio between sp$^3$ and sp$^2$ carbon atoms may alter the physical properties of 3D-graphene including the elastic properties.


  The elastic constants of 3D-graphenes were calculated by using MD method. The formula to describe the elastic constants as functions of $L$ can be obtained by theoretical analysis. In the following, we will show the detailed derivations of these functions, the validity of fitting, and the resulted scaling laws.

  \begin{figure}[htbp]
    \centering
    \includegraphics[width=1\textwidth]{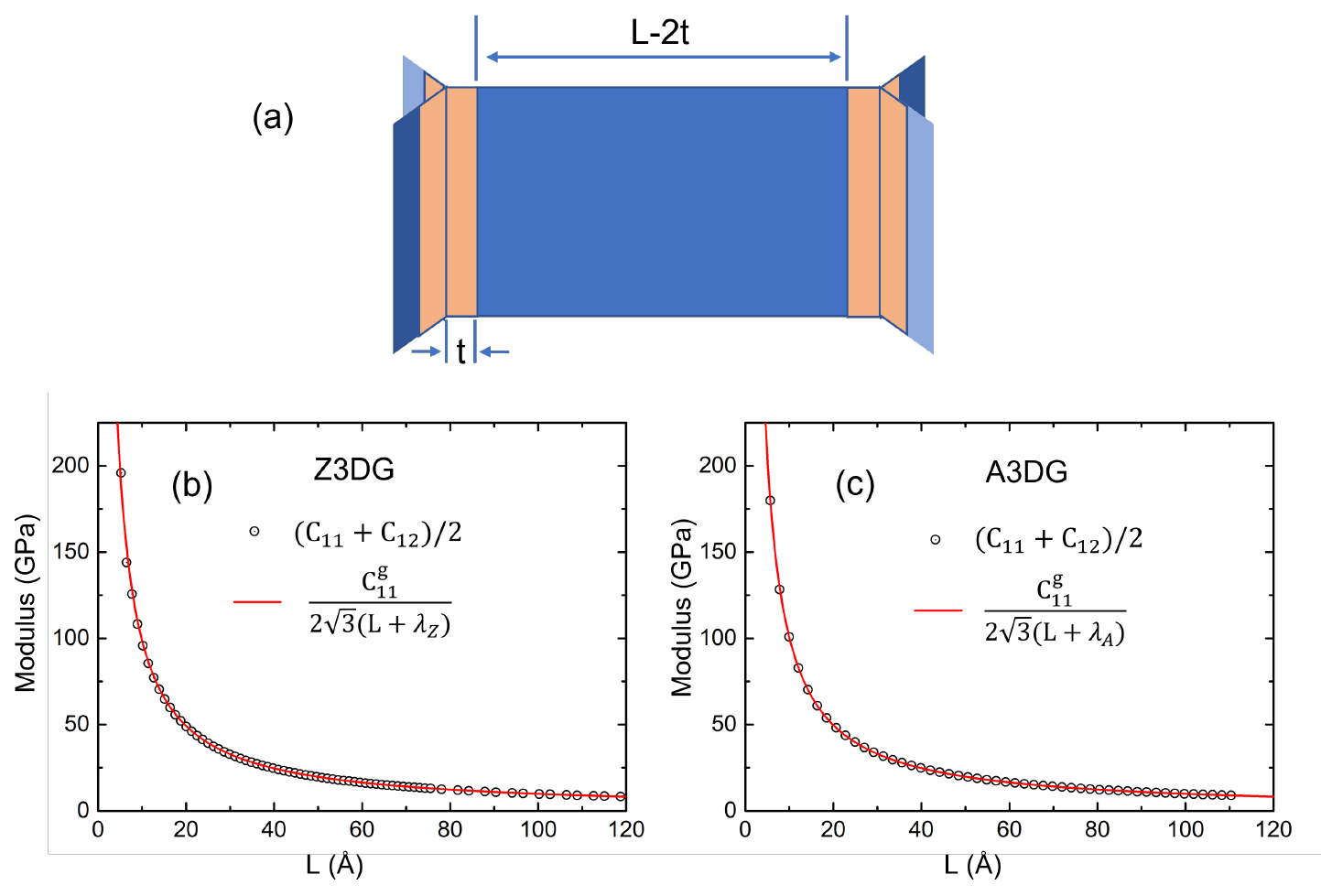}
    \caption{(a) The structure of a graphene plate with narrow stripes around the joint lines. (b) The calculated $(C_{11}+C_{12})/2$ of Z3DG and the fitted line, and (c) is that for A3DG.}
    \label{plusfig}
  \end{figure}

  As shown in Fig. \ref{3DG_structure}, there are three graphene plates in one unit cell of a 3D-graphene. They joint at two lines, one line locates inside the cell and the other line locates at the cell corner. The length of each graphene plate is $L$, the cell height is $H$. Let's consider a uniform strain in $\bm{a}$-$\bm{b}$ plane, $\bm{\epsilon}=(\epsilon,\epsilon,0,0,0,0)$, to the system. Accordingly, the strain energy is
  \begin{equation}\label{plusa}
    E=\frac{3\sqrt{3}}{2}(C_{11}+C_{12})\epsilon^2L^2H,
  \end{equation}
  where $\frac{3\sqrt{3}}{2}L^2H$ is the volume of the cell.

  In this case, the three graphene plates are equally compressed or stretched, keeping the angles between them unchanged.  Actually, the graphene plate is not totally uniform, since the area near the joint lines may be deformed. Then the areas around the joint line may have different elastic properties. This area is described effectively as a narrow stripe with width $t$, as shown in Fig. \ref{plusfig}(a), and the modulus $C_{11}^{s}$ and strain $\epsilon^s$. Now, the graphene plate without the stripes has the length $L-2t$ and the strain $\epsilon^g$. Considering the mechanical balance, the strains should meet with
  \begin{equation}\label{condition1}
    C_{11}^g\epsilon^g=C_{11}^s\epsilon^s
  \end{equation}
  And the total deformation of a graphene plate is
  \begin{equation}\label{condition2}
    L\epsilon=(L-2t)\epsilon^g+2t\epsilon^s.
  \end{equation}
  Then we have
  \begin{equation*}
    \begin{aligned}
      \epsilon^g=\frac{L\epsilon}{L+2t(C_{11}^g/C_{11}^s-1)} \\
      \epsilon^s=\frac{L\epsilon C_{11}^g/C_{11}^s}{L+2t(C_{11}^g/C_{11}^s-1)}
    \end{aligned}
  \end{equation*}
  The deformation energy for the graphene plate is
  \begin{equation*}
    \begin{aligned}
      E_{plate} & =\frac{1}{2}\frac{L^2H\epsilon^2C_{11}^g}{L+2t(C_{11}^g/C_{11}^s-1)} \\
                & =\frac{1}{2}\frac{L^2H\epsilon^2C_{11}^g}{L+\lambda}.
    \end{aligned}
  \end{equation*}
  where the parameter $\lambda$ describes the effect of narrow stripe with the positive (negative) value showing that the stripe is softer (harder) than the perfect graphene.
  Then the total strain energy in a cell is:
  \begin{equation}\label{plusb}
    E=\frac{3}{2}\frac{L^2H\epsilon^2C_{11}^g}{L+\lambda}
  \end{equation}
  Combining Eq. \ref{plusa} and \ref{plusb}, we can get the following expression:
  \begin{equation}\label{plus}
    \frac{C_{11}+C_{12}}{2}=\frac{C_{11}^g}{2\sqrt{3}(L+\lambda)}.
  \end{equation}

  The calculated results for Z3DG are shown in Fig. \ref{plusfig}(b), where the fitted line using Eq. \ref{plus} perfectly agrees with the data points. The fitted parameter is $\lambda_Z=-0.014 $\AA. The negative value shows that the stripes are harder than the perfect graphene, however, the very small value means that the effect of stripes is negligible. The results for A3DG are shown in Fig. \ref{plusfig}(c). The fitted parameter is $\lambda_A=-0.213 $\AA. The negative value also shows that the stripe is harder than the perfect graphene. It is notable that this value of A3DG is 15 times larger than that of Z3DG. This agrees with our intuition, since the carbon atoms at the stripes in Z3DG form hex-atomic rings across the joint lines, similar to graphene, on the other hand that in A3DG form much larger rings, make the properties of the stripes deviate from that of perfect graphene.

  Now we turn to the elastic constant of $C_{11}$. This component can be obtained by applying a uniaxial strain $\bm{\epsilon}=(\epsilon,0,0,0,0,0)$ to the 3D-graphene system. According to the modulus of 3D-graphene in Eq. \ref{C_hex}, the strain energy is
  \begin{equation}
    E=\frac{\epsilon^2C_{11}V}{2}=\frac{3\sqrt{3}\epsilon^2C_{11}L^2H}{4}.
  \end{equation}

  \begin{figure}[htbp]
    \centering
    \includegraphics[width=1\textwidth]{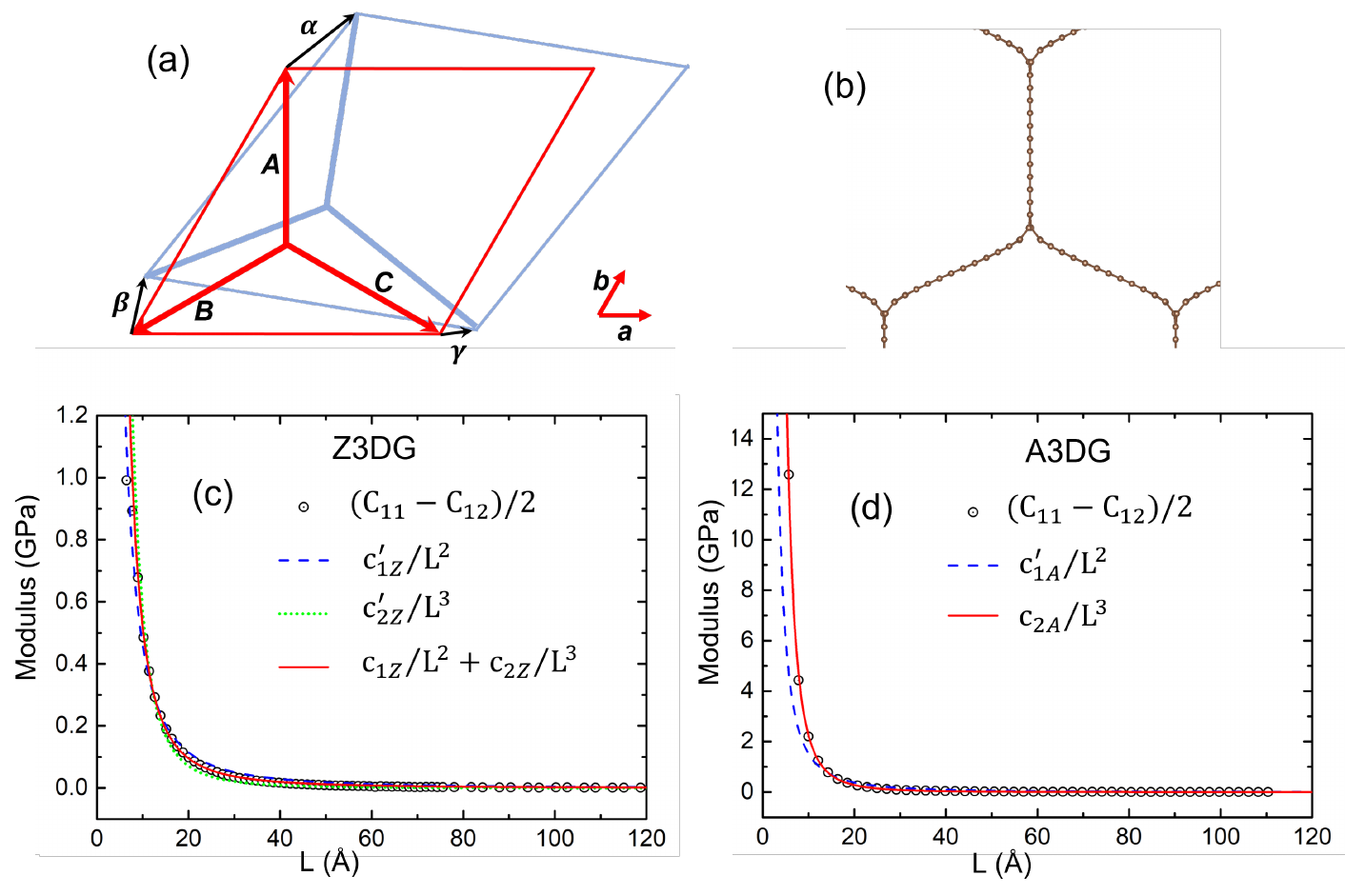}
    \caption{(a)The top-view of a general strain in $\bm{a}$-$\bm{b}$ plane applied to a 3D-graphene. (b) Top-view of a bended Z3DG-12 under strain (-0.035,0,0,0,0,0). (c) The calculated $(C_{11}-C_{12})/2$  data points and fitting lines for Z3DG, and (d) those for A3DG.}
    \label{minusfig}
  \end{figure}

  This kind of strain may alter the angles between plates, while keeping the translational symmetry along $\bm{c}$ direction. We then plot the system in the  $\bm{a}$-$\bm{b}$ plane as shown in Fig. \ref{minusfig}(a). The three graphene plates are shown as vectors $\bm{A}$, $\bm{B}$ and $\bm{C}$, along the directions from the center joint point to the three corner joint points.  We consider a general deformation, where the ends of vectors $\bm{A}$, $\bm{B}$ and $\bm{C}$ move to the new points with displacement vectors $\bm{\alpha}$, $\bm{\beta}$, and $\bm{\gamma}$, respectively.  By using the balance condition for forces at the center joint point and the linear expansion, we know that all the three graphene plates have the same amount of extension or reduction length $\Delta L$.
  \begin{equation}\label{DL}
    \Delta L = \frac{\bm{A}\cdot\bm{\alpha} + \bm{B}\cdot\bm{\beta} + \bm{C}\cdot\bm{\gamma}}{3L},
  \end{equation}
  where $L$ is the original length of each graphene plate, i.e. $L=|\bm{A}|=|\bm{B}|=|\bm{C}|$.

  For the case of  $\bm{\epsilon}=(\epsilon,0,0,0,0,0)$, we have $\Delta L=L\epsilon/2$. Thus, the strain applied to a graphene plate is $\epsilon/2$.
  Then the strain energy in a 3D-graphene cell due to the extension or reduction of the graphene plates is:

  \begin{equation*}
    E_1=\frac{3L^2\epsilon^2C_{11}^gH}{8(L+\lambda)},
  \end{equation*}

  Besides the strain of the graphene plate, the changing of solid angles at the joint lines may also introduce energy increasing. Notations of $\theta^{AB}$, $\theta^{BC}$, and $\theta^{CA}$ have been used to denote the solid angles between each two plates. Then the angle deformations can be derived at the linear order:
  \begin{equation}\label{DAngle}
    \begin{aligned}
      \Delta \theta^{AB}=-2\sqrt{3}\frac{ \bm{A}\cdot\bm{\beta} + \bm{B}\cdot\bm{\alpha} +\bm{C}\cdot\bm{\gamma} }{3L^2} \\
      \Delta \theta^{BC}=-2\sqrt{3}\frac{ \bm{A}\cdot\bm{\alpha} + \bm{B}\cdot\bm{\gamma} +\bm{C}\cdot\bm{\beta} }{3L^2} \\
      \Delta \theta^{CA}=-2\sqrt{3}\frac{ \bm{A}\cdot\bm{\gamma} + \bm{B}\cdot\bm{\beta} +\bm{C}\cdot\bm{\alpha} }{3L^2}.
    \end{aligned}
  \end{equation}
  By using these equations, the angle deformations along with $\bm{\epsilon}=(\epsilon,0,0,0,0,0)$ can be obtained, which are  $\Delta\theta^{CA}=\Delta\theta^{AB}=-\sqrt{3}\epsilon/2$, and $\Delta\theta^{BC}=\sqrt{3}\epsilon$. The energy increasing which comes from the change of the angle at harmonic approximation can be written as $KH(\Delta\theta)^2/2$ . The parameter $K$ is the stiffness of the solid angle. Then the energy originating from all the solid angle deformations at the joint lines is:
  \begin{equation*}
    E_2=\frac{9}{2}\epsilon^2KH.
  \end{equation*}

  Actually, there is a third source of energy increasing. Suppose that the solid angle did not change after applying strain $\bm{\epsilon}=(\epsilon,0,0,0,0,0)$, the graphene plates must be curved as shown in Fig. \ref{minusfig}(b). The curvature of the graphene plate introduces extra energy. The energy can be expressed by an integration of the flexural modulus of graphene times the square of the local curvature, which is proportional to $\epsilon$. If we have another 3D-graphene with twice the size of the original one, the local curvature will be reduced to half of the original value and thus be proportional to $L^{-1}$. After integration, the energy correction due to the bending of the graphene plates can be written as:
  \begin{equation*}
    E_3=c\frac{\epsilon^2H}{L}.
  \end{equation*}
  where $c$ is the integration parameter.

  Since $E=E_1+E_2+E_3$, we obtain the following modulus component
  \begin{equation*}
    \begin{aligned}
      C_{11} & =\frac{C_{11}^g}{2\sqrt{3}(L+\lambda)}+\frac{2\sqrt{3}K}{L^2}+\frac{4c}{3\sqrt{3}L^3} \\
             & =\frac{C_{11}^g}{2\sqrt{3}(L+\lambda)}+\frac{c_1}{L^2}+\frac{c_2}{L^3}.
    \end{aligned}
  \end{equation*}
  Combining this equation with Eq. \ref{plus}, we have:
  \begin{equation*}
    C_{12}=\frac{C_{11}^g}{2\sqrt{3}(L+\lambda)}-\frac{c_1}{L^2}-\frac{c_2}{L^3},
  \end{equation*}
  and
  \begin{equation}
    \frac{C_{11}-C_{12}}{2}=\frac{c_1}{L^2}+\frac{c_2}{L^3},
  \end{equation}
  where the $L^{-2}$ and $L^{-3}$ come from the solid angle deformation and the bending of graphene plates, respectively.

  The calculation results of $(C_{11}-C_{12})/2$  are shown in Fig. \ref{minusfig}(c) and (d). For Z3DG, neigher $c_1/L^2$ nor $c_2/L^3$ could fit the data points, but their combination $c_1/L^2+c_2/L^3$ fit the data points very well. Thus, both the solid angle deformation and the bending of graphene plates are crucial to $(C_{11}-C_{12})/2$ in Z3DG. The fitted parameters are $c_{1Z}=24.81$ GPa\AA$^2$ and $c_{2Z}=271.34$ GPa\AA$^3$. On the other hand, $c_1/L^2$ could not fit the data points for A3DG, while $c_2/L^3$ could fit the data points perfectly. Thus, the solid angle deformation has negligible contribution to $(C_{11}-C_{12})/2$ of A3DG. The main contribution comes from the bending of the graphene plates. The fitted parameter is $c_{2A}=2240.11$ GPa\AA$^3$, which is one order larger than $c_{2Z}$. This large difference would make the chirality of 3D-graphene significant when $L$ is small.

  \begin{figure}[htbp]
    \centering
    \includegraphics[width=1\textwidth]{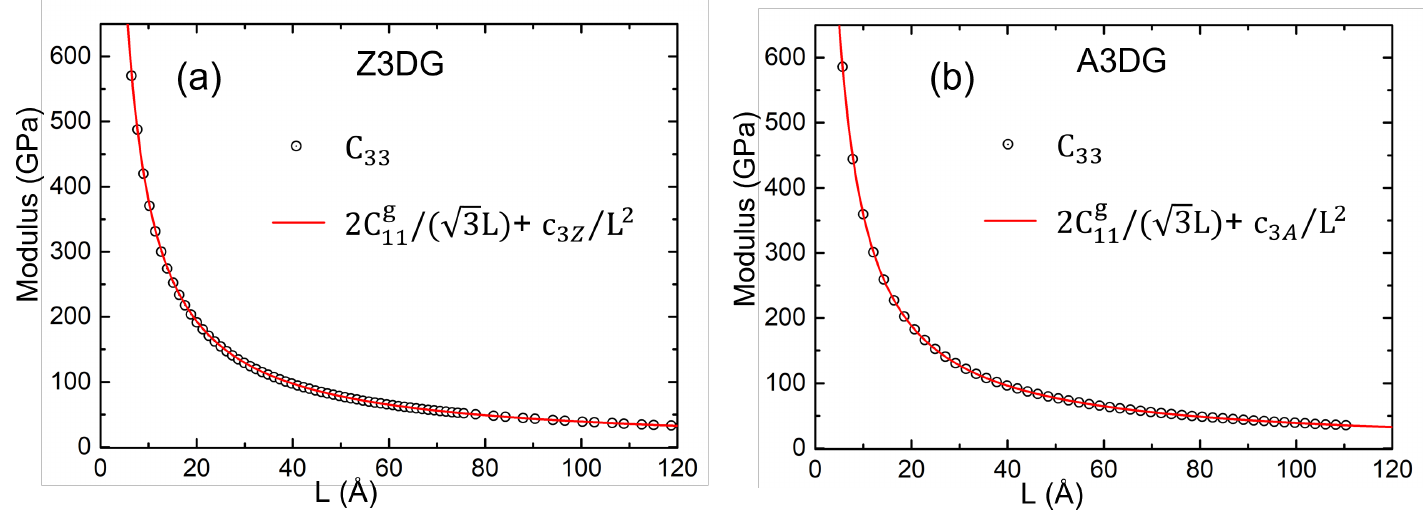}
    \caption{The calculated $C_{33}$ data points and fitting lines for (a) Z3DG and (b) A3DG.}
    \label{c33fig}
  \end{figure}

  Now, considering a uniform compression in $\bm{c}$ direction applying to the system, the corresponding strain becomes $\bm{\epsilon}=(0,0,\epsilon,0,0,0)$, and the strain energy is
  \begin{equation*}
    E=\frac{3\sqrt{3}}{4}C_{33}\epsilon^2L^2H.
  \end{equation*}
  The strain energy can also be expressed in the summation of the strain energies of each graphene plate and the strain energies of each joint line:
  \begin{equation*}
    \begin{aligned}
      E & =\frac{3}{2}C_{11}^g\epsilon^2(L-2t)H + 3C_{22}^s\epsilon^2tH + \epsilon^2kH \\
        & =\frac{3}{2}C_{11}^g\epsilon^2LH + [3(C_{22}^s-C_{22}^g)t+k]\epsilon^2H,
    \end{aligned}
  \end{equation*}
  where $k$ is the stiffness parameter of each joint line and $C_{22}^s$ is the modulus matrix element of the narrow stripes around the joint lines. Then we obtain the modulus component of 3D-graphene as:
  \begin{equation}\label{c33}
    \begin{aligned}
      C_{33} & =\frac{2C_{11}^gL}{\sqrt{3}L^2}+\frac{4[k+3(C_{22}^s-C_{22}^g)t]}{3\sqrt{3}L^2} \\
             & =\frac{2C_{11}^g}{\sqrt{3}L}+\frac{c_3}{L^2}.
    \end{aligned}
  \end{equation}
  We can see that $C_{33}$ is the combination of $L^{-1}$ and $L^{-2}$ terms. From the above expressions, we know that the ratio of $C_{33}/C_{11}$ is close to the ratio of $C_{33}/C_{12}$, and they both approach 4 at large $L$ limit.

  The calculated results for $C_{33}$ are shown in Fig. \ref{c33fig}, which demonstrates that the fitting lines from Eq. \ref{c33} perfectly fit the data points. The fitted parameters are $c_{3Z}=-1465.28$ GPa\AA$^2$ for Z3DG, and  $c_{3A}=-3473.72$ GPa\AA$^2$ for A3DG. We therefore conclude that the effects of joint lines and the narrow stripes close to them are significant for the elastic constant of $C_{33}$ in 3D-graphene.

  \begin{figure}[htbp]
    \centering
    \includegraphics[width=1\textwidth]{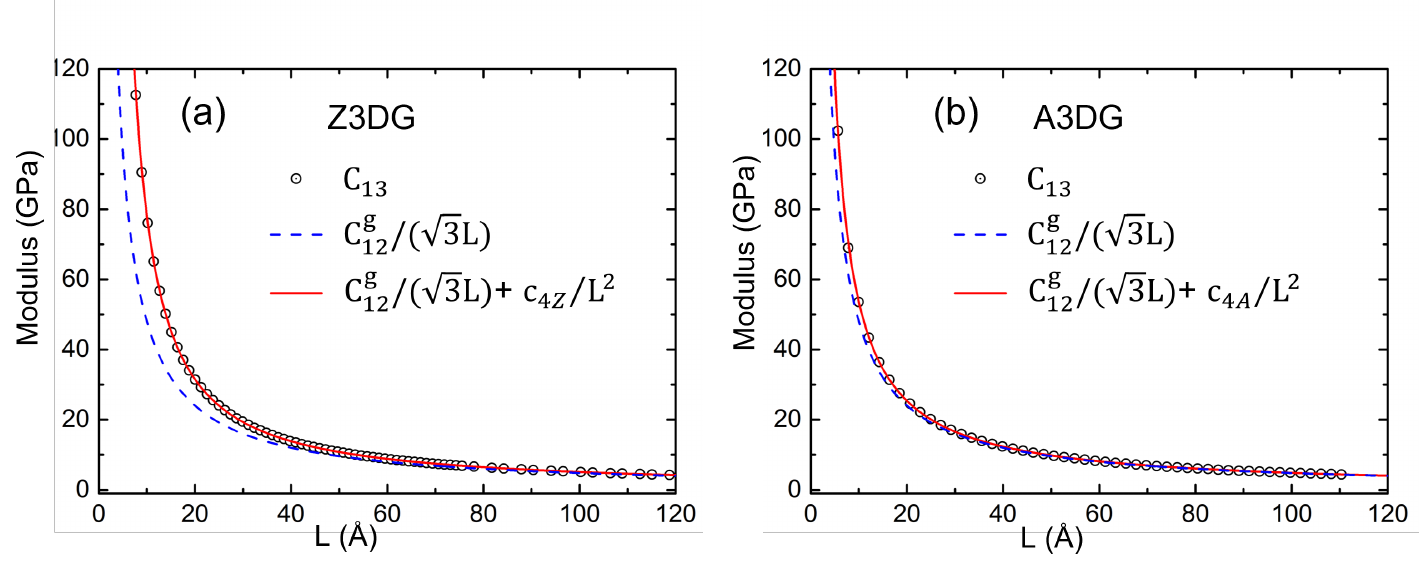}
    \caption{The calculated $C_{13}$ data points and fitting lines for (a) Z3DG and (b) A3DG.}
    \label{c13fig}
  \end{figure}

  We further apply a 3D uniform strain to the system, which has the form of $\bm{\epsilon}=(\epsilon,\epsilon,\epsilon,0,0,0)$ with the strain energy as:
  \begin{equation}\label{e13a}
    \begin{aligned}
      E=\frac{3\sqrt{3}}{4}(2C_{11}+2C_{12}+C_{33}+4C_{13})\epsilon^2L^2H.
    \end{aligned}
  \end{equation}
  The strain in a graphene plate is $\epsilon_{1}=\epsilon_{2}=\epsilon$,  and the shear strain is zero. Each graphene plate can be devided into three parts, i.e. the central perfect graphene area and the two narrow stripes near the joint lines. The strains in $\bm{c}$ direction for central part and the stripes are $\epsilon_{2}^g=\epsilon_{2}^s=\epsilon$. The strains in the other directions should satisfy the same conditions in Eq. \ref{condition1} and \ref{condition2}, which are
  \begin{equation*}
    \begin{aligned}
      \epsilon_{1}^g=\frac{L\epsilon}{L+2t(C_{11}^g/C_{11}^s-1)} \\
      \epsilon_{1}^s=\frac{L\epsilon C_{11}^g/C_{11}^s}{L+2t(C_{11}^g/C_{11}^s-1)}.
    \end{aligned}
  \end{equation*}

  The strain energies for the central perfect graphene and the stripes are
  \begin{equation*}
    \begin{aligned}
      E_1 & =\left[\frac{C_{11}^g}{2}[(\epsilon_1^g)^2+(\epsilon_2^g)^2] +C_{12}^g\epsilon_1^g\epsilon_2^g\right](L-2t)H               \\
      E_2 & =\left[\frac{C_{11}^s}{2}(\epsilon_1^s)^2+\frac{C_{22}^s}{2}(\epsilon_2^s)^2 +C_{12}^s\epsilon_1^s\epsilon_2^s\right]2tH .
    \end{aligned}
  \end{equation*}
  And the total strain energy in a cell of 3D-graphene is
  \begin{equation*}
    E=3E_1+3E_2+\epsilon^2kH,
  \end{equation*}
  where $\epsilon^2kH$ is the strain energy in the two joint lines, and $k$ is the stiffness coefficient. Combining these equations with Eq. \ref{e13a} and using the expressions of $C_{11}$, $C_{12}$, and $C_{33}$, we can get the expression for $C_{13}$ as:
  \begin{equation*}
    \begin{aligned}
      C_{13} & =\frac{C_{12}^g}{\sqrt{3}L}+\frac{2tC_{12}^sC_{11}^g}{3\sqrt{3}C_{11}^s[L+2t(\frac{C_{11}^g}{C_{11}^s}-1)]L},
    \end{aligned}
  \end{equation*}
  where the first and second terms come from graphene modulus $C_{12}^g$ and the modulus of the narrow stripe $C_{12}^s$, respectively. The contribution of the first term is much larger than that of the second term. The parameter $t$ here is a small number, thus we keep only the linear term and neglect the higher order terms of $t$:
  \begin{equation}\label{c13}
    \begin{aligned}
      C_{13} & \approx\frac{C_{12}^g}{\sqrt{3}L}+\frac{2tC_{12}^sC_{11}^g}{3\sqrt{3}C_{11}^sL^2} \\
             & =\frac{C_{12}^g}{\sqrt{3}L}+\frac{c_4}{L^2}
    \end{aligned}
  \end{equation}

  Fig. \ref{c13fig} shows the results of $C_{13}$, from which we can see that the formula with only $\frac{C_{12}^g}{\sqrt{3}L}$ term could not describe $C_{13}$ correctly. The fitting lines from Eq. \ref{c13} perfectly fits the data points. The fitted parameters are $c_{4Z}=3011.88$ GPa\AA$^2$ for Z3DG, and $c_{4A}=560.04$ GPa\AA$^2$ for A3DG.

  \begin{figure}[htbp]
    \centering
    \includegraphics[width=1\textwidth]{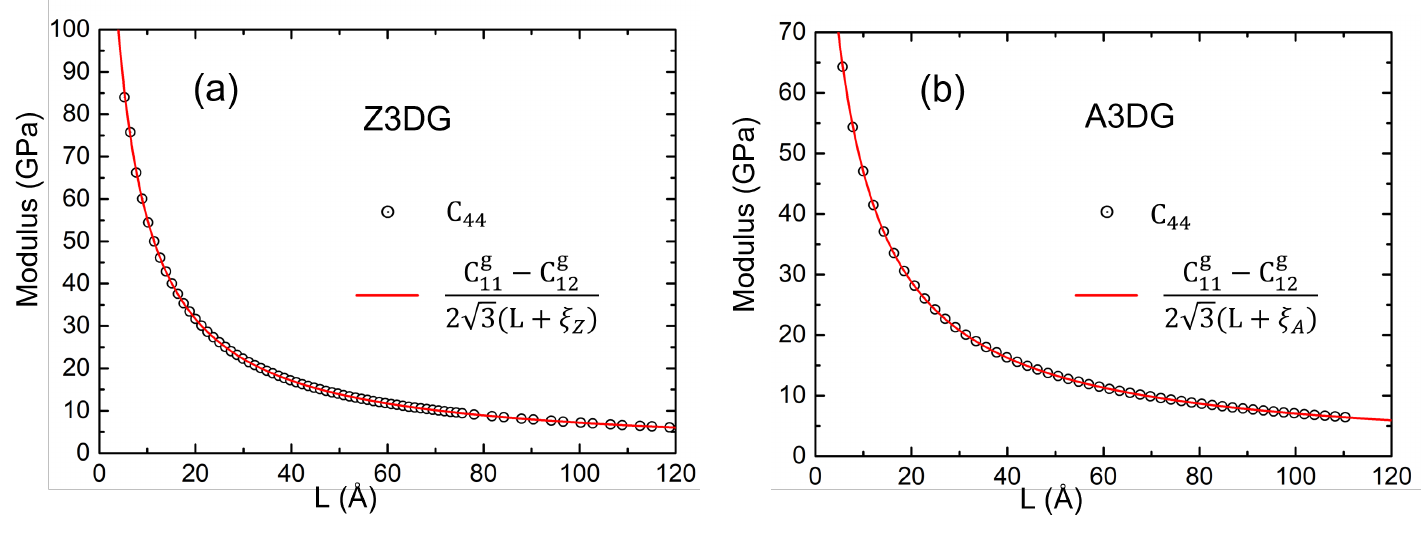}
    \caption{The calculated $C_{44}$ data points and fitting lines for (a) Z3DG and (b) A3DG.}
    \label{c44fig}
  \end{figure}

  The last independent modulus component is $C_{44}$, which can be obtained by applying a shear strain $\bm{\epsilon}=(0,0,0,0,\epsilon,0)$. The corresponding strain energy is
  \begin{equation}\label{c44a}
    E=\frac{3\sqrt{3}C_{44}\epsilon^2L^2H}{4}.
  \end{equation}
  By analyzing the distortion of each graphene plate, we know that the 2D strain components for graphene plate $\bm{A}$ are all zero, and those for graphene plates $\bm{B}$ and $\bm{C}$ are $\epsilon_{1}=\epsilon_{2}=0$, $\epsilon_{3}=\sqrt{3}\epsilon/2$. As discussed above, a graphene plate can be divided into the central graphene area and the narrow stripes attaching to the joint lines. The shear strains for these two areas can be solved as:
  \begin{equation*}
    \begin{aligned}
      \epsilon_{3}^g=\frac{\sqrt{3}\epsilon}{2}\frac{L}{L+2t(C_{33}^g/C_{33}^s-1)} \\
      \epsilon_{3}^s=\frac{\sqrt{3}\epsilon}{2}\frac{LC_{33}^g/C_{33}^s}{L+2t(C_{33}^g/C_{33}^s-1)},
    \end{aligned}
  \end{equation*}
  where $C_{33}^g=(C_{11}^g-C_{12}^g)/2$. Then the total strain energy of the two graphene plates with the nonzero shear strain is:
  \begin{equation*}
    \begin{aligned}
      E & =\frac{3}{4}\frac{C_{33}^g\epsilon^2L^2H}{L+2t(C_{33}^g/C_{33}^s-1)} \\
        & =\frac{3}{4}\frac{C_{33}^g\epsilon^2L^2H}{L+\xi}
    \end{aligned}
  \end{equation*}
  Combining this formula with Eq. \ref{c44a}, we get the following expression for $C_{44}$:
  \begin{equation}\label{c44}
    \begin{aligned}
      C_{44} & =\frac{C_{33}^g}{\sqrt{3}(L+\xi)}           \\
             & =\frac{C_{11}^g-C_{12}^g}{2\sqrt{3}(L+\xi)}
    \end{aligned}
  \end{equation}

  The calculated values for $C_{44}$ are shown in Fig. \ref{c44fig}. We can see that the fitting lines from Eq. \ref{c44} perfectly fits the data points. The fitted parameters are $\xi_Z=3.50$ \AA\, for Z3DG, and $\xi_A=5.87$ \AA\, for A3DG. The positive value means that the shear strain modulus of the narrow stripes is larger than that of an ideal graphene. Comparing the fitted parameters of $\xi$ and $\lambda$, we know that the narrow stripes near the joint lines have much more powerful influence on $C_{44}$ than to $C_{11}$ and $C_{12}$.

  \section{Conclusions} \label{Conclusions}
  In this study, we have performed a systematic investigation on the elastic properties of 3D honeycomb-like graphene structures. A hybrid neural network potential was used to numerically relax the atomic structures and calculate the elastic constants. Based on our theoretical analysis, all the five independent elastic constant components can be written as functions of the size of the graphene plate, i.e. the size of honeycomb hole in 3D-graphene. Both $(C_{11}+C_{12})/2$ and $C_{44}$ contain only one $(L+c)^{-1}$ term, where the $c$ parameter comes from the elastic response of the narrow stripes near the joint lines in 3D-graphene. The exact expressions are $\frac{C_{11}+C_{12}}{2}=\frac{C_{11}^g}{2\sqrt{3}(L+\lambda)}$ and $C_{44}=\frac{C_{11}^g-C_{12}^g}{2\sqrt{3}(L+\xi)}$. The stripes also contribute to the components of $C_{33}$ and $C_{13}$ in $L^{-2}$ terms, which can be written as $C_{33}=\frac{2C_{11}^g}{\sqrt{3}L}+\frac{c_3}{L^2}$ and $C_{13}=\frac{C_{12}^g}{\sqrt{3}L}+\frac{c_4}{L^2}$. At last, the difference between $C_{11}$ and $C_{12}$ is analyzed as $(C_{11}-C_{12})/2=\frac{c_1}{L^2}+\frac{c_2}{L^3}$, where the first term comes from the variation of the solid angles between the graphene plates and the second term comes from the bending distortion of the graphene plates. All the parameters $c_1$, $c_2$, $c_3$, $c_4$, $\lambda$ and $\xi$ are obtained by fitting the formulas to the MD data points for Z3DGs and A3DGs. The fitted parameters show that Z3DG and A3DG have different elastic properties. But their differences vanish gradually with the increasing of $L$, since all the five independent elastic constants approach to $\propto L^{-1}$, which is only relevant to the 2D elastic constants of graphene and irrelevant to the fitting parameters. Our study provides fundamental insights into the elastic properties of 3D-graphenes. The study methods and resulted scaling laws could potentially be used to other superstructures.

  ~\\
  \textbf{Acknowledgements}
  This work was supported by National Natural Science Foundation of China (Grants Nos. 12022415, 11974056, 12074271).

\end{document}